\documentclass[12pt,leqno]{article}
\usepackage{epsfig}
\usepackage{graphics,graphicx}
 \usepackage{amsmath}
 \usepackage{amssymb}
\usepackage{amsthm}
\usepackage{hyperref}

\DeclareRobustCommand{\qed}{%
  \ifmmode 
  \else \leavevmode\unskip\penalty9999 \hbox{}\nobreak\hfill
  \fi
\quad\hbox{\qedsymbol}}
\newcommand{\mathbold}[1]{\mbox{\boldmath $#1$}}

\newcommand{\pr}{\mathbold P}

\begin{document}
\begin{center}
{\large {\bf
Lasso, knockoff and Gaussian covariates: a comparison}}\\
Laurie Davies\footnote{Faculty of Mathematics, University Duisburg-Essen, 45117
   Essen, Federal Republic of Germany,
email: laurie.davies@uni-due.de}
\end{center}
\begin{abstract}
Given data ${\mathbold y}$ and $k$ covariates ${\mathbold x}_j$ one
problem in linear regression is to decide which if any of the
covariates to include when regressing the dependent variable
${\mathbold y}$ on the covariates  ${\mathbold x}_j$. In this paper
three such methods, lasso, knockoff and Gaussian covariates are compared
using simulations and real data. The Gaussian covariate method is based
on exact probabilities which are valid for all  ${\mathbold y}$ and
${\mathbold x}_j$ making it model free. Moreover the probabilities
agree with those based on the F-distribution for the standard linear
model with i.i.d. Gaussian errors. It is conceptually, mathematically and
algorithmically very simple, it is very fast and makes no use of
simulations. It outperforms lasso and knockoff in all respects by a
considerable margin.
\end{abstract}



\section{Introduction} \label{intro}
There are many papers on lasso from the first \cite{TIB96} to a very
recent one \cite{BELETS17}, a span of 21 years. As no theoretical
comparisons are made in this paper we give no further references. The
software required for the comparison is the R package glmnet which can
be downloaded from 
{\footnotesize
\begin{verbatim}
 https://CRAN.R-project.org/package=glmnet 
\end{verbatim}
}

Knockoff is much more recent. Theoretical work is to be found in \cite{CANDetal17}. The
software is obtainable from R 
{\footnotesize
\begin{verbatim}
https://cran.r-project.org/web/packages/knockoff/index.html
\end{verbatim}
}
Part of the comparison is based on the Tutorials 1 and 2  of
{\footnotesize
\begin{verbatim}
https://web.stanford.edu/group/candes/knockoffs/software/knockoff/  
\end{verbatim}
}

The present version of the Gaussian covariate method is new but it is
based on previous attempts, see \cite{DAV17}. It is described in
Section~\ref{sec:gauss_cov}. There is as yet no R package but the software is
available as an ancillary file.

The real data used in the comparison includes three of the data sets used in
\cite{DETBUH03},  colon cancer (\cite{ALOETAL99}),
leukemia (\cite{GOLETAL99}) and lymphoma (\cite{ALIETAL00}) available from\\
{\footnotesize
\begin{verbatim}
http://stat.ethz.ch/~dettling/bagboost.html
\end{verbatim}
}
For more information about the data see \cite{DETBUH02}.

The prostate cancer data is available from the {\it
  lasso2} CRAN R package 
{\footnotesize
\begin{verbatim}
 https://CRAN.R-project.org/package=lasso2
\end{verbatim}
}
and the red wine data  (\cite{COCEETAL09}) from
{\footnotesize
\begin{verbatim}
https://archive.ics.uci.edu/ml/machine-learning-databases/wine-quality/
\end{verbatim}
}

The Boston housing data and the Brownlee stack loss data are available
from the R package MASS
{\footnotesize
\begin{verbatim}
https://CRAN.R-project.org/package=MASS
\end{verbatim}
}

A further data set is the one considered in
\cite{COXBATT17} on  osteoarthritis available from the Gene Expression
Omnibus under accession number GDS5363 available from
{\footnotesize
\begin{verbatim}
https://www.ncbi.nlm.nih.gov/sites/GDSbrowser?acc=GDS5363
\end{verbatim}
}
As in \cite{COXBATT17} the males have been excluded. All the data sets
are the same but with permutations so that replicating the results
here will require exactly the same data set. The one used here is that used in
\cite{COXBATT17} and is, I think, the {\it DataSet SOFT file}.

For a short discussion of $L_1$ regression use is made of the R
package quantreg available from 
{\footnotesize
\begin{verbatim}
https://CRAN.R-project.org/package=quantreg
\end{verbatim}
}

The comparisons require the latest version of R \cite{R}. The results
of this paper can be reproduced by running the file runcomp.R. This
requires the FORTRAN file selvar.f and the R files comp.R and
selvar.R. The running time is about  eight hours.

\section{A description of the Gaussian covariate method}\label{sec:gauss_cov}

\subsection{The basic idea}
Consider data consisting of a dependent variable ${\mathbold
  y}={\mathbold y}(n)=(y_1,\ldots,y_n)$ and an explanatory variable
${\mathbold x}=(x_1,\ldots ,x_n)$. The problem is to decide if ${\mathbold x}$ is
indeed an explanatory variable for ${\mathbold y}$ in the sense that the
values of ${\mathbold y}$ can to some extent be explained by those
of ${\mathbold x}$. A standard method of deciding this is to postulate
a linear model
\begin{equation} \label{ref:model}
{\mathbold y}=\beta{\mathbold x}+\sigma {\mathbold \varepsilon} 
\end{equation}
with ${\mathbold \varepsilon}$ i.i.d. standard Gaussian noise and to test the
null hypothesis $\beta=0$. The standard test is the F-test based on
the F-statistic 
\begin{equation} \label{F}
F=(ss_y-ss_{r,x})/(ss_{r,x}/(n-1))\stackrel{D}{=}F_{1,n-1}
\end{equation}
where $ss_y=\sum_{i=1}^ny_i^2$ and $ss_{r,x}$ is the sum of the squared
residuals after regressing ${\mathbold y}$ on ${\mathbold x}$. The null
hypothesis is rejected if the P-value
\begin{equation} \label{P-val-F}
1-F_{1,n-1}(F)
\end{equation}
is less than the specified size of the test $\alpha$. 

The Gaussian covariate approach is to compare ${\mathbold x}$ with a
Gaussian covariate ${\mathbold Z}$ with i.i.d. $N(0,1)$
components. The comparison is done through the sum of the squared
residuals. The  covariate  ${\mathbold Z}$ is clearly not an explanatory covariate
so that if ${\mathbold x}$ is no better than ${\mathbold Z}$ in
respect of the sum of squared residuals it is
concluded that  ${\mathbold x}$ is also not an explanatory covariate:
the null hypothesis $\beta=0$ is replaced by the question, is 
${\mathbold Z}$ better than ${\mathbold x}$?

Denote the sum of squared residuals after regressing ${\mathbold y}$ on
${\mathbold Z}$ by $ss_{r,Z}$.  It has been shown by Lutz D\"umbgen (\cite{DAVDUE18})
that
\begin{equation} \label{B}
B=1-ss_{r,Z}/ss_y \stackrel{D}{=}B_{1/2,(n-1)/2}
\end{equation}
with P-value
\begin{equation} \label{P-val-B}
1-B_{1/2,(n-1)/2}(B)
\end{equation}
and that the two P-values are equal
\begin{equation} \label{P-val-equ}
1-B_{1/2,(n-1)/2}(B)=1-F_{1,n-1}(F).
\end{equation}
This is a remarkable result even if it has a simple proof. It is
remarkable because both P-values are exact and uniformly distributed
over $[0,1]$  but whereas the P-value on the left is valid for all
(non-zero)  ${\mathbold y}$ and ${\mathbold   x}$ that on the
right depends on the model (\ref{ref:model}). 

We need a generalization of this result also due to Lutz D\"umbgen (\cite{DAVDUE18}).
Given  ${\mathbold y}$ and $\ell$ linearly independent 
covariates ${\mathbold x}_j,j =1,\ldots, \ell$ and $\ell'-\ell \ge 1$
i.i.d. $N(0,1)$  additional random covariates we have
\begin{equation} \label{Bell}
1-ss_{\ell'}/ss_{\ell} \stackrel{D}{=}B_{(\ell'-\ell)/2,(n-\ell')/2}
\end{equation}
where $ss_{\ell}$ is the sum of squared residuals after regressing on
the  ${\mathbold x}_j,j =1,\ldots, \ell$ and  $ss_{\ell'}$ the sum of
squared residuals after regressing on all $\ell'$ covariates. The case
$\ell'=\ell+1$ is the one required for stepwise regression.

\subsection{Gaussian covariate stepwise regression} \label{meth1}
Regress ${\mathbold y}$ on ${\mathbold x}_j$ including an offset by
default, put 
\[ss_y=\sum_{i=1}^n(y_i-\text{mean}({\mathbold y}))^2\]
and denote the sum of squared residuals by $ss_j$. The best of the
${\mathbold x}_j$ is the one with the smallest  $ss_j$ given by
\[ ss_{(1)}= \min_j ss_j.\]
Denote the corresponding quantities for the Gaussian covariates by
$SS_j$ and $SS_{(1)}$ respectively. From (\ref{B})
\begin{equation}
1-SS_j/ss_y \stackrel{D}{=} B_{1/2,(n-1)/2}.
\end{equation}
As the Gaussian covariates are independent it follows that
\begin{eqnarray} \label{ref:p_value_1}
\pr(SS_{(1)} \le ss_{(1)})=1-B_{1/2,(n-1)/2}(1-ss_{(1)}/ss_y)^k.
\end{eqnarray}
If the best of the ${\mathbold x}_j$ is ${\mathbold x}_{j_1}$ the
right hand side is referred to as the P-value of ${\mathbold
  x}_{j_1}$. The smaller the P-value the more relevant  ${\mathbold
  x}_{j_1}$. This corresponds to the role of testing $\beta_{j_1}=0$
in the linear model where the smaller the P-value the more significant
the covariate ${\mathbold x}_{j_1}$.

In some applications it is useful to be less strict when selecting
variables. This may be seen as a trade-off between reducing
the number of false negatives, not selecting variables with some
explanatory value, at the risk of more false positives, selecting
variables with no explanatory value. This may be done as follows.

The random variables $B_{1/2,(n-1)/2}(1-SS_j/ss_y)$ are
i.i.d. $U(0,1)$ so that
\begin{eqnarray} 
\pr(SS_{(1)}\le ss_{(1)})&=&1-B_{1/2,(n-1)/2}(1-ss_{(1)}/ss_y)^k\nonumber\\
&=&1-B_{k,1}(B_{1/2,(n-1)/2}(1-ss_{(1)}/ss_y)) . \label{ref:p_value_1.1}
\end{eqnarray}
This can be extended to the $\nu$th order $SS_{(\nu)}$ to give
\begin{eqnarray} \label{ref:p_value_nu}
\pr(SS_{(\nu)} \le ss_{(1)})=1-B_{k-\nu+1,\nu}(B_{1/2,(n-1)/2}(1-ss_{(1)}/ss_y)).
\end{eqnarray}
Comparing $ss_{(1)}$ with $SS_{(\nu)}$ is less strict than comparing
it with $SS_{(1)}$. Although $\nu$ has a direct interpretation when an
integer this is not necessary in (\ref{ref:p_value_nu}).

To incorporate this into a stepwise procedure a cut-off
value $\alpha$ for the P-value must be specified, for example
$\alpha=0.05$. Suppose at stage $\ell$ of the procedure $\ell$
covariates have been selected. We denote the sum of the squared
residuals by $ss_{r,\ell}$ with $ss_{r,0}=ss_y$ and the set of
selected covariates by ${\mathcal S}_{\ell}$. For each $j\notin
{\mathcal S}_{\ell}$ regress ${\mathbold y}$ on the covariates in ${\mathcal
  S}_{\ell} \cup {\mathbold x}_j$ and denote the smallest sum of
squared residuals 
taken over $j$ by $ss_{\ell,(1)}$. This is now compared with the
smallest sum of squares obtainable by considering $k-\ell$ random
Gaussian covariates ${\mathbold Z}_{\kappa}, \kappa=1,\ldots,k-\ell$. These
are so to speak chosen anew for each $\ell$. This is asking the
question as to whether the remaining covariates are better than
Gaussian noise. Regressing ${\mathbold
  y}$ on the covariates in ${\mathcal  S}_{\ell} \cup {\mathbold
  Z}_{\kappa}$ gives a random sum of squared residuals
$SS_{\ell,\kappa}$. From (\ref{Bell}) we have
\begin{equation}
1-SS_{\ell,\kappa}/ss_{r,\ell} \stackrel{D}{=} \text{Beta}(1/2,(n-\ell-1)/2),
\end{equation}
and the  P-value for the best of the ${\mathbold x}_j \notin {\mathcal
  S}_{\ell}$ is given by
\begin{equation} \label{ref:p_value_2}  
\quad \pr(SS_{\ell,(1)} \le ss_{\ell,(1)})=1-\text{B}(1-
ss_{\ell,(1)}/ss_{r,\ell},1/2,(n-\ell-1)/2)^{k-\ell}. 
\end{equation}
If the P-value is greater than the cut-off value $\alpha$ the procedure
terminates. Otherwise the best covariate is included and the procedure
continues. 

The extension of the above to the $\nu$th order statistic  gives the
P-value 
\begin{eqnarray}   
\lefteqn{\qquad\pr(SS_{\ell,(\nu)} \le ss_{\ell,(1)})=}\label{ref:p_value_3}\\
&&1-B_{k-\ell+1-\nu,\nu}(B_{1/2,(n-\ell-1)/2}(1- ss_{\ell,(1)}/ss_{r,\ell}))\nonumber 
\end{eqnarray}
which is the probability that the $\nu$th best of the Gaussian
covariates is better than the best of the remaining covariates. 

The default  which includes the offset command is
{\footnotesize
\begin{verbatim}
fstepwise(y,x,alpha,kmax)
\end{verbatim}
}
The parameter {\it kmax} specifies the largest number of selected
covariates. The main reason for its inclusion without a default value
is to reduce memory size. A more experienced FORTRAN programmer may
well be able to do without it. It can also be avoided if the programme
were written in a language with a dynamic memory such as C: ``the
principle weakness of FORTRAN was and is its lack of dynamic arrays'',
\cite{HUB11} page 67. A second reason is that it enables the user to
choose a maximum number of covariates. 

Applying {\it fstepwise} to the leukemia data mentioned in
Section~\ref{intro} with $(n,k)=(72,3571)$ gives (** 1 **)
{\footnotesize
\begin{verbatim}
> fstepwise(ly.original,lx.original,0.05,10,misclass=T,time=T)[[1]]
   user  system elapsed 
  0.008   0.000   0.011 
[1,] 1182 0.0000000000 4.256962    4
[2,] 1219 0.0008577131 2.884064    3
[3,] 2888 0.0035805523 2.023725    1

\end{verbatim}
}
Here 1182, 1219 and 2888 are the selected covariates with P-values
0,  8.58e-4 and 3.58e-3 respectively. The third column gives the sum
of squared residuals after the inclusion the the covariate and the
fourth the number of misclassifications. The time
required was 0.008 seconds.

The default value $\nu=1$ can be replaced by for example $\nu=3$ by
putting {\it nu}=3 in the command to give (** 2 **)
{\footnotesize 
\begin{verbatim}
>  fstepwise(ly.original,lx.original,0.05,10,nu=3,misclass=T,time=T)[[1]]
   user  system elapsed 
  0.012   0.000   0.013 
     [,1]         [,2]     [,3] [,4]
[1,] 1182 0.000000e+00 4.256962    4
[2,] 1219 1.051452e-10 2.884064    3
[3,] 2888 7.664817e-09 2.023725    1
[4,] 1946 3.353905e-03 1.602749    1
[5,] 2102 6.026398e-04 1.242921    0
\end{verbatim}
}
Thus $\nu=3$ leads to the two additional covariates 1946 and 2102. 

\subsection{False positives and $\nu$}
The  effect of the choice of $\nu$ on the results can be estimated by
relating $\nu$ to the concept of false positives.  A false positive is
selecting a covariate which is no better than Gaussian noise.  If all
covariates are Gaussian noise then all selected covariates are
false positives. For given $(n,k,\alpha)$ the number of false positives
can be obtained using simulations which then provide a guide for real
data.  The command is 
{\footnotesize
\begin{verbatim}
fsimords(n,k,alpha,nu,kmx,nsim=100,time=T)
\end{verbatim}
}

For normal values of $\alpha$  the default value of 100 simulations is
usually sufficient so the time required is required is not very
long. Table~\ref{tab1} (** 3 **) gives the result for the
leukemia data with $(n,k,\alpha,\nu,kmx)=(72,3571,0.05,3,10)$.  The time
 was 3.49 seconds and the average number of false positives 
was 0.98. On the basis of Table~\ref{tab1}  there is no evidences
that the two additional covariates obtained with $\nu=3$ are
relevant. 
\begin{table}[h]
\begin{center}
{\footnotesize
\begin{tabular}{ccccccccccc}
\multicolumn{11}{c}{$(n,k,\alpha,\nu,kmx)=(72,3571,0.05,3,10)$}\\
0&1&2&3&4&5&6&7&8&9&mean\\
0.51&0.22&0.16&0.07&0.00&0.01&0.03&0.00&0.00&0.00&0.98\\
\quad\\
\end{tabular}
}
\caption{{\footnotesize The empirical frequencies for the number of
    false positives based on 100 simulations. }\label{tab1}} 
\end{center}
\end{table}

The definition of false costive given above is an empirical one. In
simulations based on the linear model a false positive 
is the selection of a covariate ${\mathbold x}_j$ with $\beta_j=0$. A
false negative is the omission of a covariate ${\mathbold x}_j$ with
$\beta_j \ne 0$. This is the definition we use in Tables~\ref{tab2} and
\ref{tab1}.  In the simulations of graphs in
Section~\ref{sec:graphs}  a false negative is the omission of an edge
where the true graph has an edge. A false positive is the inclusion of
an edge where the true graph has no edge. Again this is the definition
we use.

\subsection{Repeated Gaussian covariate stepwise regression} \label{repeat_1}
Once the first covariate has been chosen the stepwise procedure is
conditional on this covariate. More generally once a subset has been
chosen the next covariate to be chosen is dependent on this subset. 

To illustrate this we consider the colon cancer data  with
$(n,k)=(62,2000)$. The stepwise procedure results in (** 4 **)
{\footnotesize
\begin{verbatim}
>  fstepwise(colon.y,colon.x,1,2,misclass=T,time=T)[[1]]
   user  system elapsed 
  0.004   0.000   0.011 
     [,1]         [,2]     [,3] [,4]
[1,]  493 7.402367e-08 6.804815    9
[2,]  175 4.311166e-01 5.431871    7
\end{verbatim}
}
For any cut-off P-value $\alpha <0.43$ only one covariate is
chosen, namely 493. This does not mean that only covariate 493 is
relevant but that given 493 the remaining 1999 are in 
a sense no better than white Gaussian noise. 

If covariate 493 is eliminated and the stepwise procedure applied to
the remaining covariates again just one covariate is chosen, 377 with
a P-value 1.36e-07. The covariates 493 and 377 are highly correlated with
correlation coefficient of 0.778. This explains why 377 is no longer
considered once 493 has been included. Now 377 can be excluded and the
procedure continued in this manner until the P-value of the best of
the remaining covariates exceeds the specified cut-off value
$\alpha$. 

The value $\alpha=0.05$ results in 82 covariates being selected (** 5 **). The
time required was 0.224  seconds. The first seven are given in
Table~\ref{tab2}.  The covariates 1635  and 576 are grouped together
as indicated by the number 4. That is the two taken together give a
linear approximation to dependent variable. Similarly 1423 and 353 are
grouped together. In all the 82 covariates form 49 linear approximations.
\begin{table}[hb]
\begin{center}
\begin{tabular}{cccccccc}
&1&2&3&4&5&6&7\\
\hline
set&1&2&3&4&4&5&5\\
covariate&493&377&249&1635&576&1423&353\\
P-value&7.40e-8&1.35e-7&1.13e-6&2.28e-6&2.28e-6&2.76e-5&8.00e-4\\
ss&6.80&6.94&7.44&7.62&5.52&8.26&5.33\\
mis&9&11&8&19&5&9&6\\
\end{tabular}
\caption{The first seven of the selected 82 covariates for the colon
  cancer data. \label{tab2}} 
\end{center}
\end{table}

The default command is
{\footnotesize
\begin{verbatim}
fstepstepwise(y,x,alpha,kmax).
\end{verbatim}
}
The number of selected covariates can be reduced by specifying a
smaller P-value. Putting $\alpha=0.01$ for the colon data results in
45 covariates grouped into 32 linear approximations (** 6 **) The time
required was 0.12 seconds. The number can also be reduced by
specifying a maximum number of linear approximations using {\it nmax} or a
maximum number of covariates using {\it vmax}.  In the case of the leukemia
data running
{\footnotesize
\begin{verbatim}
fstepstepwise(ly.original,lx.original,0.05,10,time=T)
\end{verbatim}
}
results in 420 selected covariates forming 153 linear
approximations (** 7 **)  but not printed. Setting {\it nmax}=20 gives
20 linear approximations involving 62 covariates  (** 8 **)
{\footnotesize
\begin{verbatim}
fstepstepwise(ly.original,lx.original,0.05,10,nmax=20,time=T)
\end{verbatim}
}

\subsection{Misclassifications and Outliers}
\begin{figure}[h]
\begin{center}
\includegraphics[width=14cm,height=8cm,angle=0]{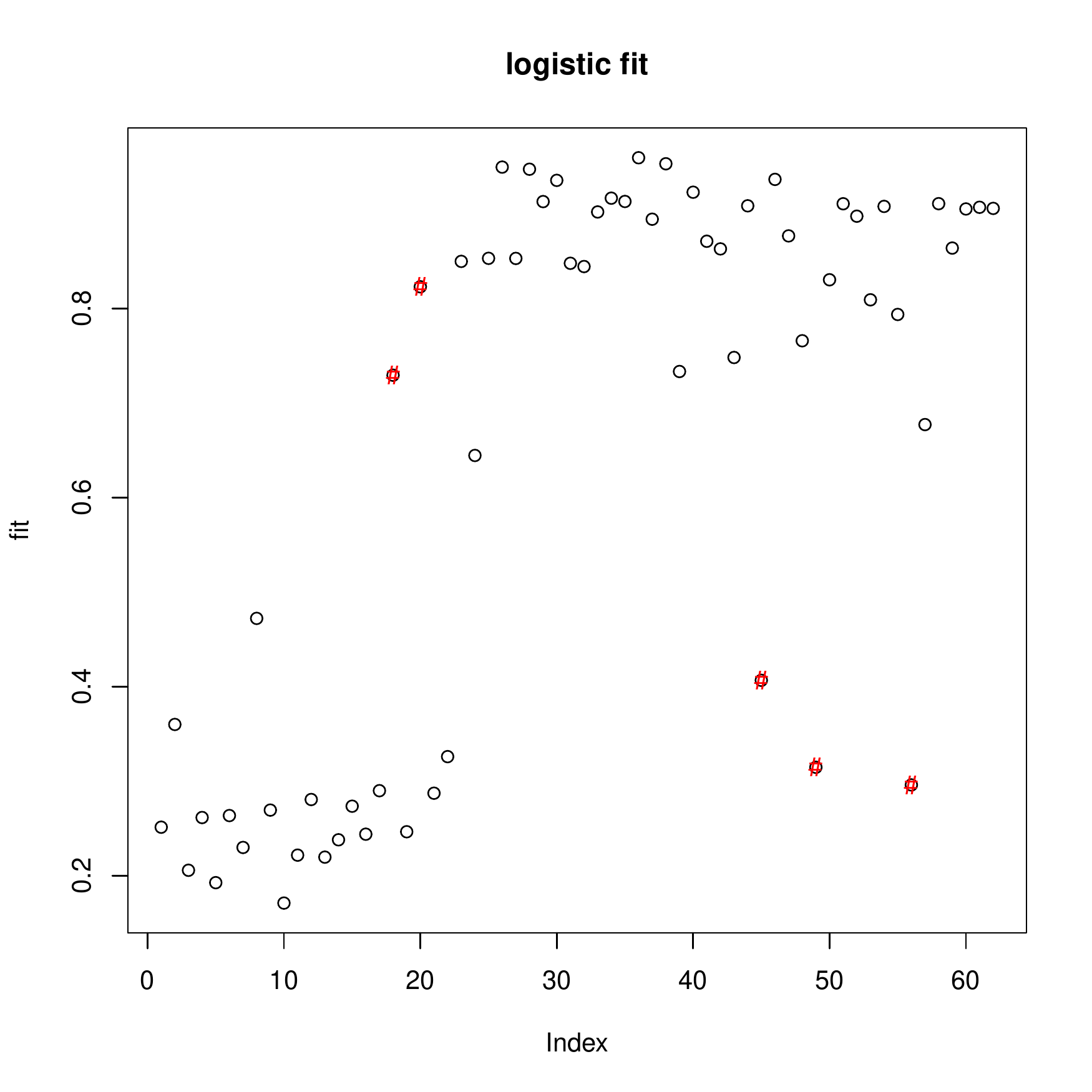}
\end{center}
\caption{The average fit of the 49 colon linear approximations using
  logistic regression. \label{fig:fit_colon}}  
\end{figure}

The repeated Gaussian stepwise procedure produces several linear
approximations each of which can be used to classify the data. the
number of misclassifications for the first five approximations for the
colon cancer data are given above. These can be combined for example,
by calculating the fit for each approximation and then taking the
average.  The result for the 49 approximations to the colon cancer data
is shown Figure~\ref{fig:fit_colon}. The least squares fit gives
similar results. There are five misclassifications shown in red and
these five are clearly outliers, very different from the other
observations. 

\subsection{$L_1$ regression}
The idea is not restricted to least squares. It can be applied to
$L_1$ regression but then simulations are necessary. We take the
Brownlee stack loss data as an example. Suppose the covariates Air
Flow and Water Temperature have been selected. Including the covariate Acid
Concentration reduces the sum of the absolute residuals
from 43.69355 to 42.08116. In a simulation with a Gaussian covariate
replacing Acid Concentration the Gaussian covariate was better in
20\% of the simulations giving a P-value of 0.204 for Acid
Concentration (** 9 **). This uses the R package quantreg. The corresponding
P-value for least squares regression is 0.344.  
{\footnotesize
\begin{verbatim}
fl1stack(stackloss,1000)
\end{verbatim}
}

For robust regression with a smooth $\psi$ function and for non-linear
regression such as logistic a chi-squared approximation is available
(see \cite{DAV17}) removing the need for simulations.

\subsection{Why Gaussian?}
The method started with the question as to whether it was possible to
judge the relevance of a covariate without assuming the model
(\ref{ref:model}). The initial data set was the Brownlee stack loss data
and the unfortunate covariate was Acid Concentration. This was
replaced by the cosine of the average daily temperature in Berlin on
the first 21 days of January 2013. The Acid Concentration won but only
just. Relying on empirical alternatives to the covariates was not a
promising option but from there it was but a short step to simulating
alternatives. Initially an attempt was made to model the alternative
covariates. If the covariate was 0-1 use the binomial, if integer
valued a Poisson etc (see page 279 of \cite{DAV14}). The results showed that the
modelling was not worth it. They were essentially the same when one
used Gaussian alternatives.

In the case of one Gaussian covariate ${\mathbold Z}=(Z_1,\ldots,Z_n)$
the sum of squared residuals is 
\begin{equation} \label{equ:ressq}
\frac{(\sum_{i=1}^n y_iZ_i)^2}{\sum_{i=1}^nZ_i^2}\stackrel{D}{\approx}
\frac{\sum_{i=1}^n y_i^2}{n} \chi^2_1.
\end{equation}
If the $Z_i$ are not Gaussian but say Bernoulli $\pm 1$ then
(\ref{equ:ressq}) will still hold asymptotically but this will require
conditions on ${\mathbold y}$. More generally if the $Z_i$ have finite
variance then subject to conditions on the $y_i$ (\ref{equ:ressq})
will hold. In this case there seems to be no reason not to use the
exact result for Gaussian covariates. 

We start from $\sum_{i=1}^ny_i^2$ so when calculating the percentage
reduction in the sum of squares due to regression on the $Z_i$ the
expression $\sum_{i=1}^ny_i^2$ cancels out just as $\sigma^2$ cancels
out in the F-test. This is why the calculated probabilities hold for
all ${\mathbold y}$. If the $Z_i$ are replaced by i.i.d. Cauchy random
variables $C_i$ then (\ref{equ:ressq}) becomes
\[\frac{(\sum_{i=1}^n y_iC_i)^2}{\sum_{i=1}^nC_i^2}=\left(\sum_{i=1}^n
  \vert y_i\vert\right)^2\frac{{\tilde C}^2}{\sum_{i=1}^nC_i^2}\]
where ${\tilde C}$ is a standard Cauchy random variable. There is no
cancelling out and the distribution will depend on ${\mathbold y}$
even asymptotically.  Moreover $(\sum_{i=1}^n
  \vert y_i\vert)^2$ is larger than $\sum_{i=1}^ny_i^2$
indicating that Cauchy random variables are less exacting than
Gaussian random variables. 

\subsection{A summary}
This completes the description of the Gaussian covariate method. It is
extremely simple. There is no mention of regression parameters $\beta$
or the variance $\sigma^2$ of (\ref{ref:model}). This contrasts with
the treatment of lasso in \cite{BELETS17} where all the values of the
tuning parameter $\lambda$ considered involve $\sigma$.  Indeed the
estimation of $\sigma$ is one of the main problems with lasso as many
optimality results for the choice of $\lambda$ depend on the value of
$\sigma$. The linear approximations, or models if the reader wishes,
provided by  the repeated stepwise procedure are as they stand. They
take explicitly into account the number of available covariates, they
do not over fit and there is no need for any form of post selection
analysis. 

As an example we take the colon data. All the linear
approximations provided by the repeated stepwise procedure have either
one or two covariates. The first one consists only of the covariate
493. If we use logistic regression using this covariate alone there
are nine misclassifications. Replace the remaining 1999 covariates by
Gaussian noise and choose the first three covariates. One of these is
always 493, the two remaining are Gausssian noise. Simulations  show
that in about 2.8\% of the cases there are zero misclassifications and
in about 3.3\% just one misclassification. This indicates that you can
get overfitting with just three covariates.

\section{Comparison of lasso, knockoff and the Gaussian covariate
  procedure}
\subsection{Problems with interpretation}
The comparisons given here are purely empirical. It is possible to
prove theoretical results on the Gaussian covariate method as a first
attempt in  \cite{DAV17} shows but this will not be pursued
further. The comparisons are given in detail and it should be 
possible to repeat them using the software available as an auxiliary
file. The version of lasso to be
used is the cross-validation option {\it cv.glmnet} provided in the R
package glmnet.

One problem when comparing lasso and knockoff with the Gaussian
covariate method is how to interprete the outputs of lasso and
knockoff. One applications of lasso to the colon cancer data with the
binomial family option  resulted in the four covariates (**10 **)
\[249,  377,  493,  625.\]
The time required was 0.64 seconds.

With $fdr=0.9$ and the binomial family option Knockoff selected 10 covariates (** 11 **)
\[14,  249,  493,  576,  625,  792, 1360, 1473, 1679, 1843.\]
The time required was about 65 minutes.

If these are interpreted as models then for a sample size of $n=62$
both over fit to such an extent to make them unacceptable. For this
data set you can get overfitting with just four  Gaussian covariates.

                                   ** 9.5 ** overfit
The first four Gaussian covariates are chosen and the number of
misclassifications based on the logit model determined. In 18\% of the
 cases this was less than 4. Overfitting occurs already with
three covariates. Reasonable approximations should have either one or two
covariates.  This is the case for the  49 linear approximations of **
5**.

 A second problem is that the number of covariates selected by lasso
 and knockoff varies from application to application. This is
 particularly pronounced for knockoff.  Ten applications of knockoff
 resulted in from 5 to 12 covariates being chosen.  (** 20 **). 

 A third  problem is the amount of time required. Again this is particularly
 a problem for knockoff which required 65 minutes to select the model
 based on the 10 covariates given above (** 11 **). The repeated
 Gaussian method  required 0.17 seconds to provide 49 perfectly
 reasonable approximations.

The simulations described below use a models so that it is possible to
give the number of false positives and negatives. For the red wine and
Boston housing data the goal is to give a good linear approximation to
the data. In such cases we use the output of lasso and knockoff and
make no attempt at a post choice analysis

For the gene expression data sets, the colon data, the
leukemia data,   the prostate cancer data  and the
osteoarthritis data  the dependent variable is zero-one depending on
whether the person has or has not the medical condition under
investigation. Very often logistic regression is used to analyse such
data but here we use least squares which is much easier and in terms of
stability much better.  We point out however the the Gaussian covariate
approach can be adopted to non-linear regression such as logistic
regression.  In the lymphoma data there are two different medical
conditions plus the control group. The two conditions can be analysed
separately but here we treat the data set as a whole.

For these data sets it can be argued that the problem is not to
specify a set of linear approximations but to list those covariates
which are significantly associated with the dependent
covariate. This is argued in \cite{COXBATT17}, \cite{DETBUH02},
\cite{DETBUH03} and, we think, in \cite{CANDetal17}. Nevertheless even
if this is the main goal it may still be of interest to compare linear
approximations. The sum of the squared residuals is part of the output
of {\it fstepwise} and {\it fstepstepwise} but this can be augmented by
setting {\it misclass =T} which then gives the number of
misclassifications. This is based on the least squares residuals but a
logistic regression can easily be used once the covariates are
given. The difference is small.

\subsection{Post selection analysis for lasso and knockoff}
The Gaussian covariate method associates a P-value with each chosen
covariate enabling the statistician to form a judgment about the
relevance of the covariate. Lasso and knockoff provide so to
speak naked covariates without any indication of their individual relevance.
We now describe two methods for associating a P-value with each
individual covariate. 

Two illustrate the first method we we take the 4 covariates given
above specified by one application of lasso (** 10 **) and the 10
covariates resulting from one application of knockoff (** 11 **).

We consider all subsets  of size three or less of the covariates
specified by the method used (lasso, knockoff). For each such subset
${\mathcal S}$ we calculate the P-value of each covariate $i \in
{\mathcal S}$  as given by (\ref{ref:p_value_3}) with $\nu=1$
\[1-\text{pbeta}(\text{pbeta}(1-ss_{\mathcal S}/ss_i,0.5,(n-s-1)/2),k-s+1,1)\]
where $n$ is the sample size, $k$ the number of covariates, $ns$ the
size of ${\mathcal S}$, $ss_{\mathcal S}$ the sum of squared residuals when
regressing the dependent variate on all covariates in ${\mathcal S}$ and
$ss_i$ the sum of squared residuals when the covariate $i$ is
omitted. The offset is included by default. This compares the
covariate $i$ with the best of $k-s+1$ Gaussian covariates. 

To reduce overfitting we require that the P-value of each covariate in
${\mathcal S}$ is less than a specified threshold $alpha_1$ The
value we use is $alpha_1=0.05$. The final P-value for the covariate
$i$ is its minimum P-value over all subsets which contain it and satisfy
the threshold condition. We also restrict attention to those
covariates with P-value less than a second threshold value {\it alpha}
which here is also taken to be 0.05.

The default command is
{\footnotesize
\begin{verbatim}
fpval1(y,x,ind,alpha,alpha1)
\end{verbatim}
}
where {\it(y,x)} is the data, {\it ind} the chosen covariates, {\it
  alpha} and {\it alpha1}  the threshold values.

The results for the lasso (** 10 **) and knockoff (** 11 **) selections
are given by (** 12 **) and (** 13 **) respectively. The output is
\[(i, p(i),i2,i3,ss,nms)\]
where $i$ is the covariate, $p(i)$ the P-value of $i$, $i2$ and $i3$
are the further covariates in the subset ${\mathcal S}_i$ for which the P-value of $i$
is smallest. A $0$ denotes their absence. The final two items are the
sum of the squared residuals $ss$ and, if {\it misclass=T},
the number of misclassifications based on ${\mathcal S}_i$. Using {\it
  fpval1} all of the lasso covariates are included and 8 of the 10
konockoff covariates.

A variation on this method is to take a subset ${\mathcal S}$ of size
$\kappa$ and then to choose the best covariate $j$ which minimizes the
sum of squared residuals based on ${\mathcal S}\cup \{j\}$. This will
in general introduce new covariates which were not in the original
selection.  The default command is
{\footnotesize
\begin{verbatim}
fpval(y,x,ind,alpha,alpha1)
\end{verbatim}
}
The additional covariates are give a minus sign if they are in the initial choice ${\mathcal S}$.
 For {\it fpval} the results are again all lasso covariates and again
 8 knockoff covariates but with 1360 instead of 1473.

The second method can be used when the set of selected covariates  is
too large for the first method. It uses the Gaussian
covariate method but restricts the choice of covariates to the set of
selected covariates and adjusts the P-value to take into account that these
covariates were selected from a larger set. The command is
{\footnotesize
\begin{verbatim}
fstepwise(y,x[,ind],alpha,kmax,ek=k)
\end{verbatim}
}
or
{\footnotesize
\begin{verbatim}
fstepstepwise(y,x[,ind],alpha,kmax,ek=k)
\end{verbatim}
}
depending on whether just one or all relevant linear approximations are
required. Here {\it ind} is the initial selection of covariates and
{\it k} is the number of covariates from which this selection was made.

Another method of measuring the relevance of covariates is to
calculate the cross-validation error.  This is given by
{\footnotesize
\begin{verbatim}
tmpcv<-cv.glmnet(colon.x[,ind],colon.y,family="binomial")
print(min(tmpcv$cvm))
\end{verbatim}
}
where $ind$ denotes the covariates. Using the first three covariates
from the stepwise method 493, 175 and 1909 gives a cross validation
error of 0.39,  ** 13.1 **. The four lasso covariates one of  0.58,  ** 13.2 **and the 10
knockoff covariates one of 0.46, ** 13.3 **.

\subsection{Tutorials 1 and 2} \label{sec.tut}
The simulations of Table~\ref{tab2} are based on the Tutorials 1 and 2
respectively of with the parameters as given there. 

The default choice $\nu=1$ avoids false positives
possibly at the cost of false negatives. Using this value of $\nu$ in
Tutorial 1 (** 14 **) results in on average 15 covariates being chosen with no
false positives. Putting $\nu=5$ in  results in 53
covariates being chosen with three false positives (Table~\ref{tab3})
which agrees well with the expected number 2.1 of Table~\ref{tab4}. 
Thus compared with $\nu=$ the choice $\nu=5$ increases the number of
selected covariates from 14 to 49 of which according to
Table~\ref{tab3} about 2 may be false positives. 
Putting $\nu=10$ results in 62 being chosen of which about seven  are
false positives (Table~\ref{tab3}) . Compared to $\nu=5$ there is an
increase of about 13 in the number of selected covariates of which
about 4 may be false positives. Similar calculations apply to
Tutorial 2 (**15 **). 

It follows from Table~\ref{tab3} that lasso selects on average
approximately 140 covariates in Tutorial 1 and 100 in Tutorial 2 which
is somewhat excessive. 

In terms of the sum of false positives and false negatives knockoff
and the choices $\nu=5$ and $\nu=10$ are comparable.  The default
value of the false discovery rate fdr is 0.1 but in Tutorial 2 it is set to 0.2. If the
default value 0.1 is used the false positive and negative values
become 2.48 and 44.6 respectively. Also in Tutorial 2 the covariance matrix
{\it Sigma} was used to construct the knockoff variables. This seems to
improve the performance but only slightly. If the knockoff filter is
used as in Tutorial 1 the average numbers of false positives and negatives become
6.32 and 36.72 respectively.

The main difference between knockoff and the Gaussian covariate method is
the computing time. In Tutorial 1 knockoff is 20 times slower that the
Gaussian method with $\nu=10$ and 50 times slower in Tutorial
2.
60.1 15.0 11.5
\begin{table}
\begin{center}
{\footnotesize
\begin{tabular}{cccccccc}
&\multicolumn{3}{c}{Tutorial 1}&&\multicolumn{3}{c}{Tutorial 2}\\
method&fp&fn&time&&fp&fn&time\\
\hline
lasso&82.4&0.52&7.89&&60.1&15.0&11.5\\
knockoff&5.58&10.0&63.9&&7.00&35.1&53.9\\
$\nu=1$&0.00&46.2&0.25&&0.00&56.5&0.04\\
$\nu=5$&3.36&11.6&2.29&&2.78&42.5&0.44\\
$\nu=10$&7.02&5.82&3.33&&7.00&35.4&0.94\\
\hline
\quad
\end{tabular}
}
\caption{{\footnotesize Comparison of lasso,  knockoff and Gaussian
    covariates with $\alpha=0.05$ based on Tutorials 1 and 2 (** 14
    **) and (** 15 **)}. \label{tab3}}
\end{center}
\end{table}

\begin{table}[h]
\begin{center}
{\footnotesize
\begin{tabular}{ccccccccccccc}
\multicolumn{13}{c}{$(n,k,\alpha)=(1000,1000,0.05)$}\\
&0&1&2&3&4&5&6&7&8&9&10&mean\\
$\nu=5$&0.13&0.29&0.20&0.20&0.12&0.02&0.03&0.00&0.01&0.00&0.00&2.13\\
$\nu=10$&0.02&0.02&0.04&0.08&0.15&0.11&0.17&0.20&0.09&0.04&0.08&5.84\\
\quad\\
\end{tabular}
}
\caption{{\footnotesize The empirical frequencies and the expected
    number of false positives based on 100 simulations (** 16 **). }\label{tab4}} 
\end{center}
\end{table}

The commands for Tutorial 1 and Tutorial 2 are as follows: 
{\footnotesize
\begin{verbatim}
ftut(1,1000,1000,60,4.5,0.1,0.05,50)
ftut(2,1000,1000,60,7.5,0.2,0.05,50)
\end{verbatim}
}

\subsection{Red wine data}
For the  red wine data with $(n,k)=(1599,11)$ the
dependent variable is variable 12 and gives subjective evaluations
of the quality of the wine with integer values between three and eight.

Ten applications of lasso gave either 4 or 6  selected
covariates union was $\{1 ,2 , 5,  7, 10, 11\}$.
Ten  applications of knockoff gave between 6 and 11 so that all
covariates were selected.selected  (** 17 **).  The Gaussian method with
cut-off P-value $\alpha=0.05$ gives six covariates which are in order
alcohol, volatile-acidity, sulphates,  total-sulfur-dioxide,
chlorides, pH (see Table~5 of  \cite{LOKTAYTIB214}) (** 18 **). The
repeated stepwise method gives  two more linear approximations with 4
and a single covariate  (** 19 **). 
{\footnotesize
\begin{verbatim}
fstepwise(redwine[,12],redwine[,1:11],0.05,11,misclass=T)
fstepstepwise(redwine[,12],redwine[,1:11],0.05,11,misclass=T)
\end{verbatim}
}

\subsection{Cancer data}

\subsubsection{Colon data}
The size of colon cancer data is $(n,k)=(62,2000)$.

Five application of lasso resulted in  either 4 or 5 covariates with
union $\{ 249,  377,  493,  625, 1772\}$. 4,  5,  25 and 29
variables giving 32 variables in all . Five  applications of the
knockoff filter with fdr =0.5 resulted  in 5, 12, 8, 5 and  7
variables giving 13 variables in all (** 20 **).  
The repeated Gaussian covariate method for the colon cancer data with
$alpha=0.05$was treated in  Section~ \ref{repeat_1}. It results in 82
covariates forming 49 linear approximations (** 5 **).  The computing
times were two seconds for lasso,  41 minutes for
knockoff and 0.01 seconds for the Gaussian covariate method. 
All the lasso covariates and 10 of the 13 knockoff covariates are
included in the Gaussian list (** 21 **).

The post selection P-values for the lasso with {\it alpha=alpha1=0.05}
using {\it fpval1}  and {\it fpval} are given in (**
22 **), those for the knockoff covariates in (** 23 **). Of 
the 13 knockoff covariates two have P-values greater than 0.05 and so
are not included. ** 23.5 ** gives the corresponding P-values for
Gaussian covariate selection.

\subsubsection{Leukemia data}
The size of the leukemia data is $(n,k)=(72,3571)$. 

The knockoff procedure with fdr=0.5  resulted in  31 covariates.  The
time required was five hours 20 minutes. In view of this no further
analysis was carried out as it would require of the order of a day's
computing time.

The lasso is much faster requiring 0.3 seconds on average. Five
applications of lasso resulted in 14, 12, 14, 16 and 14 covariates giving 17
covariates in all  (** 24 **).   The lasso P-values based on {\it fpval1}
are given by (** 25 **) with {\it alpha1=alpha=0.05}  result in 16
of the 17 covariates. The repeated Gaussian method (** 26 **) with $alpha=0.05$
gives 420 covariates forming 153 linear approximations. These include
15 of the 17 lasso covariates. The Gaussian procedure with {\it
  nmax=20} results in 62 covariates which include 15 of the original
17  lasso covariates (** 26 **)

\subsubsection{Prostate cancer}
The size of the prostate data $(n,k)=(102,6033)$ which suggest a very long
time using the knockoff filter. It will not be considered.

Lasso was applied 5 times with 24, 14, 14, 7  and 23  covariates being
selected giving 24 in all  (** 27 **). The P-value
procedure with {\it alpha=alpha1=0.05} reduces this to 18 (** 28 **). The
repeated Gaussian method with $alpha=0.05$ resulted in 278 covariates
and 118 linear approximations. Putting {\it nmax=20} gives 52
covariates (** 29 **). The 278 include 16 of the original 24 lasso
covariates, the 52 include 16 of the 18 {\it fpval1} lasso covariates (** 30  **) 

\subsubsection{Lymphpoma} 
The lymphoma data are the only ones where the dependent variable takes
on three values, 0, 1 and 2. The size is $(n,k)=(62, 4026)$. 

Five application of lasso resulted in 40, 41, 38, 44 and 44 covariates
giving 46 in all (** 31 **). The P-value method with $alpha=alpha1=0.05$
includes all of them (** 32 **). 

The repeated Gaussian method results in 1603 covariates forming 512
linear approximations.  It contains all the lasso covariates. Putting
{\it nmax =20} results in 78 covariates which include 28 of the lasso
covariates (** 33 **).

\subsubsection{Osteoarthritis}
The size of the osteoarthritis data is $(n,k)=(129, 48802)$.
This proved too large for the knockoff filter which required 17.7 GB
of memory.

Lasso give 10 covariates in all (** 34 **). The P-value method (** 35
**) included all of them.

The osteoarthritis data set was analysed in (\cite{COXBATT17}). The
authors selected the following 17 covariates. 
{\footnotesize
\begin{verbatim}
7235 11643 25125 25470 25744 27642 27920 29679 33385 
36409 37443 44276 45991 46771 48415 48433 48549
\end{verbatim}
}
The P-value method with reduces this to 16 
covariates (** 36 **).

The repeated Gaussian method yields  317 covariates and 107 linear
models. These include all the lasso and 11 of the Cox-Battey
covariates. Putting {\it nmax=20} reduces this to 62 covariates which
include all the lasso and 3 of the Cox-Battey covariates (** 37 **).

\subsection{Boston housing data and interactions}
For the Boston housing data $(n,k)=(506,13)$. Allowing for interactions
of order up to and including seven increases the number of covariates
from 13 to 77520 giving $(n,k)=(506,77520)$ (** 39 **).  The command is
{\footnotesize
\begin{verbatim}
fgeninter(x,ord)
\end{verbatim}
}

Five applications of lasso resulted in 39, 44, 53, 53 and 43 covariates
giving in all 61 (** 40 **). The time required was eight minutes. To
choose a reasonable linear approximation from some subset of these
covariates we apply the Gaussian stepwise procedure just to these
covariates but set the effective sample {\it ek} size to 77520 (** 41
**). The results of a linear regression based on these covariates is
given in (** 42 **). The sum of squared residuals is 6493. The
covariates are decomposed in (** 43 **). 

The Gaussian stepwise selection is given in (** 44 **). The time
required was about two seconds.  This is followed by a linear
regression (** 45 **) with sum of squared residuals 5576. The
decomposition of the covariates (** 46 **).

 The last three steps are repeated for interactions of degree
eight or less (** 47 **), (** 48 **), (** 49 **) and (** 50 **). It
may be noted that the sum of squared residuals for regression based on
the first three interactions, 441, 197063 and 197166,  is 10417. This is less
than the standard regression on all 13 initial covariates with sum of
squared residuals 11078

\subsection{Graphs} \label{sec:graphs}
Given covariates ${\mathbold x}_j, j=1,\ldots ,k$ a graph is
calculated as follows. Each ${\mathbold x}_j$ is regressed on the
remaining covariates and connected to those covariates found to be
relevant

As an example we take ${\mathbold x}$ to be a set of covariates generated
in Tutorial 1 of
{\footnotesize
\begin{verbatim}
https://web.stanford.edu/group/candes/knockoffs/software/knockoff/  
\end{verbatim}
}
The graph is given by a bidiagonal matrix with 1 on the main diagonal
and 0.25 on the side diagonals ** 51 **.

One application of lasso  ** 52 ** resulted in 4 false negative and 82 false
positives. The time requires was about 150 minutes.

The Gaussian covariate method was  applied with the cut-off P-value
$\alpha$ was replaced by $\alpha/k$. This resulted in 1 false negative and no
false positives ** 53 **. The time required was 13 seconds.

In \cite{MEIBUE06} lasso was used to calculate graphs for large$k$. 
In a simulation in Section~4 of that paper with
$(n,k)=(600,1000)$ the method resulted in two false positives and 638
false negatives.  The description of the generation of the graph of Figure~1
is incorrect and has been altered to reproduce graphs with about 1800
edges. One application of lasso gave 571 false positives and 13 false
negatives. The time taken was 46 minutes, ** 54 **

For the same data the Gaussian covariate procedure with $\alpha=0.05$ 
and $nu=1$ resulted in zero false positives and 118 false
negatives, ** 55 **. The time required was 10 seconds.
Putting $nu=2$ resulted in 8 false positives and 9 false negatives in
12 seconds. ** 56 **. Simulations suggest on average about 6 false
positives. ** 57 **.

Graphs can also be constructed for real data sets. The colon cancer
data with $(n,k)=(62,2000)$. Lasso requires 10 minutes and produces a
graph with  23322 edges.  ** 59 **.
The Gaussian covariate method with $\alpha=0.05$ gives a graph with
1634 edges in 2.8 seconds.  The first ten edges are given ** 59.5 **

The repeated Gaussian covariate method with $\alpha=0.05$ gives a
graph with 24475 edges in 44 seconds  **  60 **

Putting $\alpha=1e-7$ gives a graph with 1521 edges. The first 10 are
given ** 61 **

It is seen that the repeated method picks up many more highly
significant dependencies than the single method so the recommendation
would seem to be to use the repeated method with a smaller value of
$\alpha$.

Finally for the  osteoarthritis data with $(n,k)=(129,48802)$ lasso
requires 11 seconds for one node resulting in an estimated
11*48802/(24*3600)=6.2 days for the whole graph. The Gaussian
covariate method requires about 0.24 seconds for each node giving an
estimated time of 3 hours 15 minutes for the whole graph.  A specially
written FORTRAN programme gives a graph with 38986 edges in 45 minutes.
** 62 **

This data set shows the limitations of the recommendation made just
above. Regress the first covariate on the remaining 44801 
where $\alpha$ has been divided by the number of variables as in the
default version of fgraphst. This results in 4009 selected covariates.
If a graph were to be constructed the first covariate alone would be
connected to 4009 other covariates. The calculated P-values for the
first 152 covariates are zero. Such a graph would be much too large
to be useful. 

In this particular case one can make use of the problem and restrict
the construction of the graph to those covariates chosen in the first
step. We take the 74 covariates selected by the Gaussian covariate
method and apply the repeated Gaussian procedure to 
calculate a graph with $\alpha=1e-6$. It has 289 edges. The first
covariate 939 is connected with  18 of the 62 covariates. ** 63 **.

\section{Acknowledgment}
We thank Oliver Maclaren for comments on earlier versions.

\bibliographystyle{apalike}
\bibliography{literature}
\end{document}